\documentclass{article}
\usepackage{spconf,amsmath,graphicx}
\usepackage[numbers,sort&compress]{natbib}
\usepackage[OT1]{fontenc} 

\title{Occupancy-map-based rate distortion optimization for video-based point cloud compression}
\name{Li Li$^{\ast}$ \qquad Zhu Li$^{\ast}$ \qquad Shan Liu$^{\dag}$ \qquad Houqiang Li$^{\star}$ }

\address{$^{\ast}$ University of Missouri Kansas City   \\
         $^{\dag}$ Tencent America \\
         $^{\star}$ University of Science and Technology of China }
\begin{document}
\maketitle
\begin{abstract}
The state-of-the-art video-based point cloud compression scheme projects the 3D point cloud to 2D patch by patch and organizes the patches into frames to compress them using the efficient video compression scheme.
Such a scheme shows a good trade-off between the number of points projected and the video continuity to utilize the video compression scheme.
However, some unoccupied pixels between different patches are compressed using almost the same quality with the occupied pixels, which will lead to the waste of lots of bits since the unoccupied pixels are useless for the reconstructed point cloud.
In this paper, we propose to consider only the rate instead of the rate distortion cost for the unoccupied pixels during the rate distortion optimization process.
The proposed scheme can be applied to both the geometry and attribute frames.
The experimental results show that the proposed algorithm can achieve an average of $11.9\%$ and $15.4\%$ bitrate savings for the geometry and attribute, respectively.
\end{abstract}
\section{Introduction}
\label{sec:intro}
% introduce point cloud and point cloud compression
Point cloud is a set of 3D points with each point associated with some attributes such as color, reflection, and so on.
As point cloud has the capability to render an object or scene, the point cloud can be used in many scenarios \cite{Tulvan2016} such as 3D immersive telepresence \cite{Fuchs2014}, virtual reality, and some other applications.
However, the large size for the point cloud especially the dynamic point cloud (DPC) is preventing the usage of such a promising media format.
For example, for a DPC with $300$ frames, each frame with one million points, each point with $30$ and $24$ bits to represent the geometry and attribute, the original size can be as large as $2$GBytes.
Therefore, there is an urgent need to compress the DPC efficiently.

% point cloud compression is difficult
In addition to its large size, the isolated points also make the point cloud difficult to be compressed due to the irregular patterns.
The current methods used for DPC compression can be roughly divided into two groups: the 3D-based method and the video-based method.
As its name implies, the 3D-based method compresses the point cloud in the original domain.
Usually, the geometry of the first frame will be compressed using the octree tree \cite{schnabel2006octree} or binary tree combined with the quadtree \cite{kathariya2018scalable}.
Some transforms such as Graph Fourier Transform (GFT) \cite{Zhang2014} and region-adaptive hierarchical transform (RAHT) \cite{Queiroz2016} or layer based prediction \cite{Khaled2017} are used to compress the attributes.
For the inter frames, the key is to exploit the intercorrelations among various frames of the point cloud.
For example, Queiroz and Chou \cite{Queiroz2017} proposed to divide the frames into multiple cubes and perform the translational motion model based motion estimation to find the corresponding cube in the previously coded frame.
Mekuria \emph{et al.} \cite{Mekuria2017} further introduced the iterative closest points instead of the translational motion model to improve the coding efficiency.
However, the 3D-based schemes are unable to fully exploit the correlations among various views, which make the compression ratio unsatisfiable.

% current solution and problems
The video-based method which projects the 3D point cloud into 2D space to fully utilize the intercorrelations using the mature video compression schemes such as High Efficiency Video Coding (HEVC) \cite{Sullivan2012}.
For example, Schwarz \emph{et al.} \cite{Schwarz2017} and He \emph{et al.} \cite{He2017} proposed to use cylinder or cube projection to project the point cloud to a 2D video, and the 2D video is compressed using HEVC.
However, such kind of methods may lose lots of points due to the points occlusion in 3D space.
To solve this problem, Mammou \emph{et al.} \cite{Khaled2017} \cite{schwarz2018emerging} proposed to project the point cloud to 2D video patch by patch to increase the number of points projected.
The projected video is then compressed using HEVC and wins the competition of the MEPG call for proposal for the dynamic point cloud compression \cite{Marius2017}.
Such a video-based method will be called as patch-based point cloud compression in the following sections.
However, many unoccupied pixels of the projected video are not beneficial for the quality of the reconstructed video, which will lead to a serious waste of the bits if they are encoded in the same quality with the occupied pixels.
The unoccupied pixels are even not continuous in the temporal domain, which may lead to even more bits waste.

% contribution
In this paper, we propose to use the occupancy-map-based rate distortion optimization (RDO) to solve this problem.
When the whole or just a part of the block is unoccupied, only the rate instead of the rate distortion (RD) cost of the unoccupied pixels is taken into consideration in the RDO process.
Such a scheme is applied not only to the mode decision process but also to the sample adaptive offset (SAO) \cite{fu2012sample} process.
Under the proposed scheme, the unoccupied pixels will be coded with very bad quality to save the bitrate, while the occupied pixels are encoded in the same quality with the original scheme to guarantee the quality of the reconstructed point cloud.
As far as we can see, this is the first work trying to design specific RDO algorithm to improve the efficiency of the point cloud compression.

% paper organization
This paper is organized as follows.
We will give a brief introduction of the patch-based point cloud compression method and the motivation of the proposed algorithm.
We will describe the proposed occupancy-map-based RDO algorithm in Section~\ref{sec:proposed}.
In Section~\ref{sec:experiments}, we will show the detailed experimental results.
Section~\ref{sec:conclusion} concludes the whole paper.

\section{Motivation}
\label{sec:motivation}

Under the patch-based point cloud compression method, the 3D point cloud is first projected to its bounding box to generate several patches.
Then all the patches are packed properly to form the geometry and attribute frames.
The patch information including occupancy map, patch position in the 2D frame, and some other information is first transmitted to the decoder, and then the geometry and attribute video are encoded using HEVC.
In the lossy point cloud compression, the occupancy map is losslessly encoded as $4\times4$ blocks to obtain a better trade-off between the bitrate and accuracy.

\begin{figure}[tb]
\centering
\begin{minipage}[b]{0.4\linewidth}
  \centerline{\includegraphics[width=4.0cm]{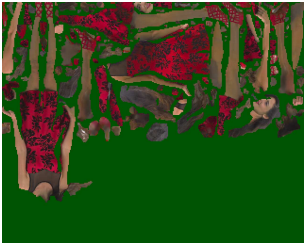}}
\end{minipage}
\caption{Frame with unoccupied pixels set as $0$ from redAndBlack, POC 16.}
\label{fig:occupancy}
\end{figure}

Fig.~\ref{fig:occupancy} gives a typical frame with unoccupied pixels set as $0$.
As we can obviously see from Fig.~\ref{fig:occupancy} that there are many unoccupied pixels among different patches.
Except those unoccupied pixels in the bottom of the frame which are easy to be encoded using intra prediction, the unoccupied pixels among different patches may cost lots of bits.
As these unoccupied pixels will have no influences on the quality of the reconstructed point cloud, the bits spent are totally useless which may seriously degrade the compression performance.
In addition, under the patch-based point cloud compression scheme, the occupancy map can be encoded before the geometry and attribute videos, which makes it possible for us to use such kind of information to improve the geometry and attribute compression.
Therefore, we propose to use an occupancy-map-based RDO to solve the problem of the unoccupied pixels in this paper.

\section{Proposed algorithm}
\label{sec:proposed}
In the default encoder of HEVC reference software, the encoding parameters $P$ including mode, motion and residue of a block are determined by the RD cost $J$.
\begin{equation}
\label{Eq:RDO_org}
\min_P J = \sum\limits_{i=1}^{N} D_i + \lambda R
\end{equation}
where $D_i$ is the distortion of each pixel, $R$ is the bitrate of a block, and $N$ is the number of pixels in a block.
During the RDO process, the $D_i$ is the square difference between the original pixel and the reconstructed pixel for a specified position $i$. 
However, such an optimization problem treats the occupied pixels and the unoccupied pixels with equal importance, which may degrade the performance of the point cloud compression significantly.

In the proposed solution, a distortion mask is added to (\ref{fig:occupancy}) to handle the unoccupied pixels in the RDO process.
\begin{equation}
\label{Eq:RDO_mask}
\min_P J = \sum\limits_{i=1}^{N} D_i \times M_i + \lambda R
\end{equation}
where $M_i$ is $1$ when the current pixel is occupied and $M_i$ is $0$ when the current pixel is unoccupied.
We then apply such a formula in the intra mode decision, inter mode decision, and SAO process to obtain good performance.

For intra prediction, the mode decision process can be divided into $3$ stages: rough intra mode selection, intra mode decision, intra residue quadtree decision.
In the first stage, the sum of the absolute transformed difference (SATD) between the original signal and the prediction signal is used as the distortion $D$.
The bitrate of the Luma intra prediction direction is used as the bitrate $R$.
In this stage, the aim is to obtain several candidates which can provide more accurate prediction blocks.
As the residue bits are not considered in this stage, we cannot consider only part of the pixels to calculate the distortion as the distortion can be considered as an indicator of the residue bitrate.
Without considering the total residue bitrate, we may select the prediction mode candidates with too many residue bits which is not beneficial for the RD performance.
In the second and third stages, as the total bitrate is taken into consideration, we apply (\ref{Eq:RDO_mask}) to both the mode decision and residue quadtree processes.

\begin{figure}[tb]
\centering
\begin{minipage}[b]{0.48\linewidth}
  \centerline{\includegraphics[width=4.0cm]{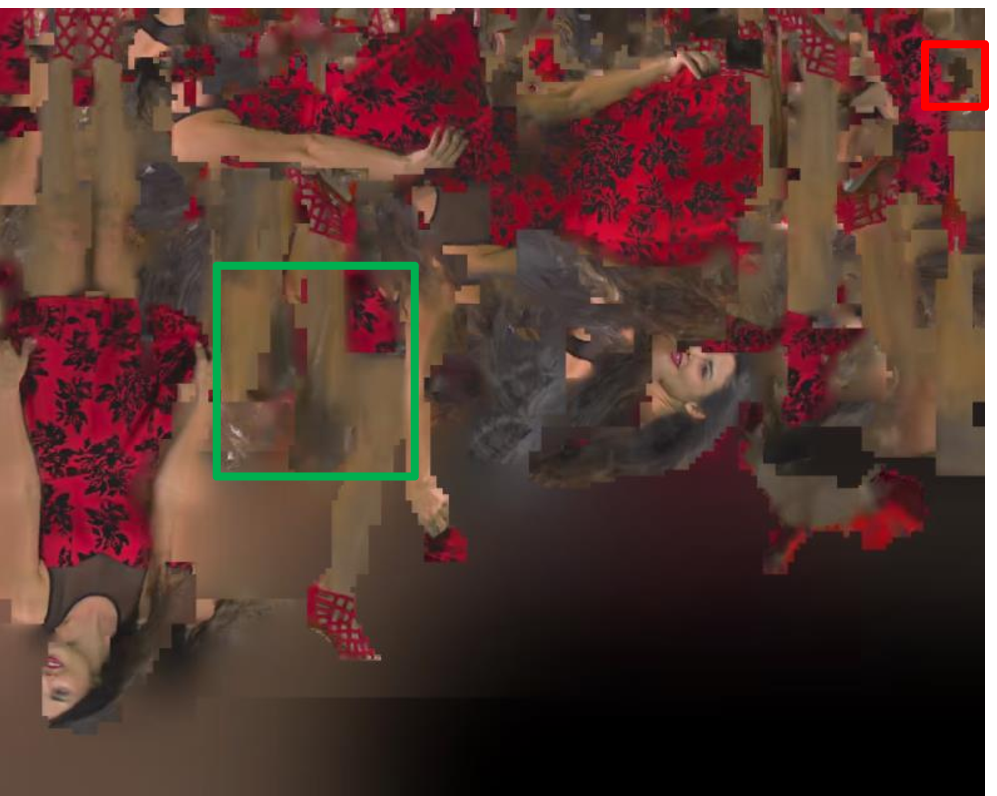}}
  \centerline{(a) Occupancy-map-based RDO}\medskip
\end{minipage}
\hfill
\begin{minipage}[b]{0.48\linewidth}
  \centerline{\includegraphics[width=4.0cm]{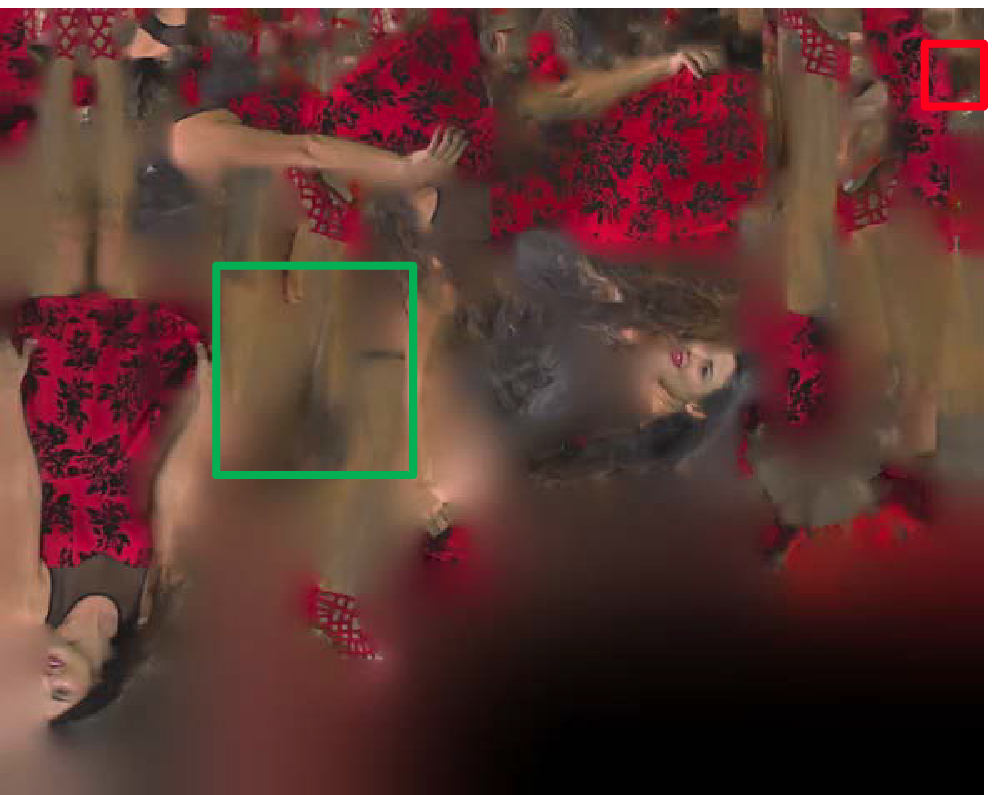}}
  \centerline{(b) Original RDO}\medskip
\end{minipage}
\caption{Reconstructed frame comparison from redAndBlack, POC 16. }
\label{fig:RDO_comparison}
\end{figure}

For inter prediction, the mode decision can be divided into two modes: $2N\times2N$ merge/skip and other modes including $2N\times2N$ advanced motion vector prediction mode and all the other partitions.
For the $2N\times2N$ merge/skip mode, all the merge candidates are considered using the full RDO, which calculates the sum of the squared difference (SSD) between the original signal and the reconstructed signal as the distortion and the overall bits including the header and residue as the bitrate.
Therefore, Eq.~(\ref{Eq:RDO_mask}) is applied directly for the $2N\times2N$ merge/skip mode.
For the other modes, the RDO process can be roughly divided into two stages: the integer/fractional motion estimation (ME) or merge to determine the prediction and the residue determination.
In the integer ME process, the sum of absolute difference (SAD) between the original signal and prediction signal is used as the prediction, and the bits of the motion vector (MV) are used as the bitrate.
In the fractional ME or merge process, the SATD between the original signal and prediction signal is used as the prediction, and the bits of the MV or merge index are used as the bitrate.
As in both cases, the residue bitrate is not taken into consideration, we cannot consider only part of the pixels to calculate the distortion.
In the residue determination, as the full RDO is used, Eq.~(\ref{Eq:RDO_mask}) is used in the RDO process.

In the SAO process, the offsets between the original pixels and the reconstructed pixels are first counted for different kinds of edge offsets and the band offset.
Then for each CTU, we will select from these kinds of offsets using RDO to determine the SAO type.
From the above descriptions, we know that the offset for different types of SAO is the key to influence the RDO process.
However, during the mode decision process, the occupancy-map based RDO encodes some of the unoccupied pixels with very small bitrate and bad quality.
If we statistic the offsets among all the pixels, the unoccupied pixels will take the majority of the offsets since their distortions are much larger.
In this way, the offset for each type of SAO will be quite inaccurate for the occupied pixels but quite accurate for the unoccupied pixels, which may lead to serious bits cost to recover the unoccupied pixels while degrading the quality of the occupied pixels during the RDO process for each CTU.
Therefore, we do not explicitly change the RDO process of SAO but change the statistic process to get the offset.
We will calculate the offsets according to only the occupied pixels between the original signal and the reconstructed signal.
The RDO process will then automatically choose to use the offset for the occupied blocks while not to use the offset for the unoccupied blocks.

Fig.~\ref{fig:RDO_comparison} shows the reconstructed frame comparison between the occupancy-map-based RDO and the original RDO.
From Fig.~\ref{fig:RDO_comparison} and Fig.~\ref{fig:occupancy}, we can see that most occupied pixels are kept the same, which means the reconstructed point clouds under the proposed algorithm will be with almost the same quality as the original RDO.
From the red rectangles in Fig.~\ref{fig:RDO_comparison}, we can see that some of the occupied pixels can be reconstructed with better quality since we can focus on finding a more suitable MV for the occupied pixels.
From Fig.~\ref{fig:RDO_comparison} and Fig.~\ref{fig:occupancy}, we can also see that many unoccupied pixels are encoded with very bad quality from the green rectangles.
Some of them are just copied directly from the temporal co-located position to save the bitrate.
Since they will not have any influences on the reconstructed quality of the point cloud, saving some bitrate is beneficial for the overall point cloud compression.

\section{Experimental results}
\label{sec:experiments}
We implement the proposed algorithm in the video-based point cloud compression (V-PCC) reference software \cite{PCCC2} and the corresponding HEVC reference software \cite{HM-16.18-SCC} to compare with the V-PCC anchor.
We test all the five DPCs defined in the V-PCC common test condition \cite{Sebastian2018}.
We test both the lossy geometry lossy attribute in all intra and random access cases to show the benefits of the proposed algorithm in intra and inter cases, respectively.
We test all the DPCs for $32$ frames, which can be a good representative of the whole sequence, to save some encoding time.
We also follow the rate points defined in the V-PCC CTC and the BD-rate is employed for comparisons of various algorithms.

For the BD-rate of the geometry, we report the results with both the point-to-point error (D1) and point-to-plane error (D2) as the quality metrics.
For the attributes, the results for Luma, Cb, and Cr are provided.
Both the encoder and decoder complexities of the V-PCC and HEVC reference software are provided since we made modifications in both the V-PCC and HEVC reference software.

\begin{table}[tp]
\begin{center}
\small
\caption{\label{tab::inter}%
Performance of the proposed occupancy-map-based RDO compared with the V-PCC anchor in random access case}
{
\begin{tabular}{c|cc|ccc}
\hline
Test         &  \multicolumn{2}{c|}{Geom.BD-Rate} & \multicolumn{3}{c}{Attr.BD-Rate}     \\
point cloud  &  D1    & D2     &    Luma   &  Cb   &    Cr                     \\
\hline
Loot        &   --16.3\%  & --16.4\%   &  --24.3\%   & --18.2\%  &  --19.3\%          \\
RAB         &   --6.6\%   & --7.2\%    &  --12.2\%   & --9.8\%   &  --12.3\%        \\
Solider     &   --15.8\%  & --16.0\%   &  --16.8\%   & --9.4\%   &  --9.0\%          \\
Queen       &   --13.4\%  & --13.2\%   &  --15.7\%   & --11.2\%  &  --10.5\%        \\
LD          &   --7.5\%   & --7.8\%    &  --7.9\%    & --7.7\%   &  --7.2\%        \\ 
\hline
Avg.        &   --11.9\%  & --12.1\%   &  --15.4\%   & --11.3\%  &  --11.7\%        \\
\hline
Enc. self &                          \multicolumn{5}{c}{101\%}                                           \\
Dec. self &                          \multicolumn{5}{c}{88\%}                                           \\
Enc. child &                         \multicolumn{5}{c}{99\%}                                            \\
Dec. child &                         \multicolumn{5}{c}{88\%}                                           \\
\hline
\end{tabular}}
\end{center}
\end{table}

\begin{figure}[tb]
\centering
\begin{minipage}[b]{0.58\linewidth}
  \centerline{\includegraphics[width=5.8cm]{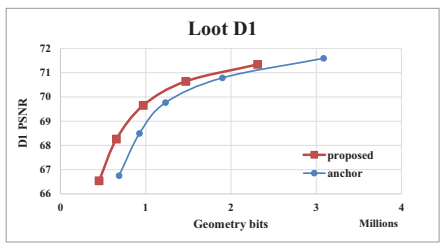}}
  \centerline{(a) Loot D1}\medskip
\end{minipage}
\hfill
\begin{minipage}[b]{0.58\linewidth}
  \centerline{\includegraphics[width=5.8cm]{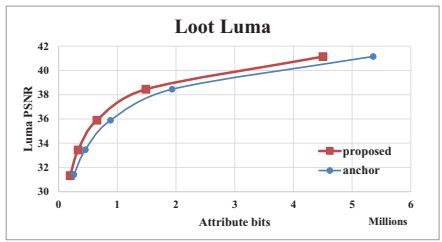}}
  \centerline{(b) Loot Luma}\medskip
\end{minipage}
\caption{Typical R-D curves for geometry and attribute. }
\label{fig:RDcurve}
\end{figure}

Table~\ref{tab::inter} shows the performance of the proposed occupancy-map-based RDO compared with the V-PCC anchor in random access case.
From table~\ref{tab::inter}, we can see that the proposed occupancy-map-based RDO can lead to an average of $10.9\%$ and $11.3\%$ performance improvements compared with the V-PCC anchor for the D1 and D2, respectively.
We can also see that the proposed algorithm can save averagely $15.0\%$, $9.2\%$, and $10.4\%$ bitrates for the Luma, Cb and Cr, respectively.
The proposed algorithm obviously demonstrates that the proposed algorithm can lead to significant bitrate savings by ignoring the distortions of the unoccupied pixels.
Sometimes, the proposed algorithm can also lead to slightly better predictions for the occupied pixels since only the distortions of the occupied pixels are considered when searching through the candidates of the merge candidate list.
We also show two example R-D curves in random access case for both the geometry and attribute as shown in Fig.~\ref{fig:RDcurve}.
From Fig.~\ref{fig:RDcurve}, we can also obviously see that the proposed algorithm can bring very significant bitrate savings for both the geometry and attribute.

The proposed algorithm will not have significant influences on the V-PCC encoder and decoder themselves since we only output the occupancy map information from the encoder.
However, for the HEVC encoder, the proposed algorithm can lead to significant time savings for the encoder and decoder since the motion estimation and motion compensation processes are both significantly simplified.
For the encoder, there is no need to search over too many merge candidates and MVs especially for the unoccupied pixels thus the encoding complexity can be reduced.
For the decoder, more blocks will choose zero motion vectors, which lead to fewer interpolation operations and save the decoding complexity.

\begin{table}[tp]
\begin{center}
\small
\caption{\label{tab::intra}%
Performance of the proposed occupancy-map-based RDO compared with the V-PCC anchor in all intra case}
{
\begin{tabular}{c|cc|ccc}
\hline
Test         &  \multicolumn{2}{c|}{Geom.BD-Rate} & \multicolumn{3}{c}{Attr.BD-Rate}     \\
point cloud  &  D1    & D2     &    Luma   &  Cb   &    Cr                     \\
\hline
Loot        &   --3.4\%  & --3.5\%   &  --1.4\%   & --0.5\%   &  --0.9\%          \\
RAB         &   --2.7\%  & --3.1\%   &  --1.1\%   & --0.9\%   &  --1.4\%        \\
Solider     &   --2.9\%  & --3.2\%   &  --1.1\%   & 0.7\%     &  1.2\%          \\
Queen       &   --2.6\%  & --2.5\%   &  --1.2\%   & --1.3\%   &  --2.0\%        \\
LD          &   --2.7\%  & --2.9\%   &  --0.7\%   & --0.7\%   &  --0.8\%        \\ 
\hline
Avg.        &   --2.9\%  & --3.0\%   &  --1.1\%   & --0.5\%   &  --0.8\%        \\
\hline
Enc. self &                          \multicolumn{5}{c}{101\%}                                          \\
Dec. self &                          \multicolumn{5}{c}{94\%}                                           \\
Enc. child &                         \multicolumn{5}{c}{98\%}                                           \\
Dec. child &                         \multicolumn{5}{c}{88\%}                                           \\
\hline
\end{tabular}}
\end{center}
\end{table}

Table~\ref{tab::intra} shows the performance improvement of the proposed occupancy-map-based RDO compared with the V-PCC anchor in all intra case.
From Table~\ref{tab::intra}, we can see that the proposed occupancy-map-based RDO algorithm can achieve an average of $1.9\%$ and $2.1\%$ performance improvements for the D1 and D2, respectively.
The proposed algorithm can achieve averagely $0.9\%$, $0.2\%$, and $0.3\%$ performance improvements for the Luma, Cb and Cr, respectively.
The performance of the proposed algorithm is quite obvious but not as significant as the inter case.
Since we cannot change the rough intra prediction direction process as mentioned in Section~\ref{sec:proposed}, what we have mainly changed is only the final intra prediction direction and residue quadtree.
Compared with the inter case where we can influence the MV in the merge candidates and the partition determination, the changes of the proposed algorithm in intra are much smaller.
In the all intra case, the proposed algorithm also leads to almost no complexity changes to the V-PCC software in both the encoder and decoder while fewer complexities for the HEVC reference software.

\section{Conclusion}
\label{sec:conclusion}
In this paper, to prevent the wasting of the bitrate on the unoccupied pixels in the video-based point cloud compression (V-PCC) scheme, we propose to consider only the rate instead of the rate distortion cost for the unoccupied pixels during the rate distortion optimization process.
The proposed idea is applied to intra prediction, inter prediction and sample adaptive offset processes.
The proposed algorithm is implemented in the V-PCC and the corresponding High Efficiency Video Coding reference softwares.
The experimental results show that the proposed algorithm can lead to averagely $11.9\%$ and $15.4\%$ bitrate savings for the geometry and attribute, respectively.
In the future, we will further consider the blocks in the patch boundary with part of blocks occupied while the other part of blocks unoccupied to further improve the performance.

% -------------------------------------------------------------------------

\end{document}